\documentstyle[moriond,epsfig,11pt]{article}
\def\ep{\epsilon}
\def\De{\Delta}

\def\ra{\rightarrow}

\def\th{\theta}
\def\Th{\Theta}

\def\bk{{\bf k}}

\def\de{\delta}
\def\la{\lambda}
\def\be{\begin{equation}}
\def\ee{\end{equation}}
\def\bea{\begin{eqnarray}}
\def\eea{\end{eqnarray}}
\def\MM{{\cal M}}
\def\SS{{\cal S}}
\def\GG{{\cal G}}
\newcommand{\lsim}{\stackrel{\textstyle <}{\sim}}
\newcommand{\gsim}{\stackrel{\textstyle >}{\sim}}

\begin{document}

\title{COSMOLOGICAL STRUCURE FORMATION WITH TOPOLOGICAL DEFECTS}
\author{Ruth DURRER}
\address{D\'epartement de Physique Th\'eorique, Universit\'e de
Gen\`eve, 24 Quai E. Ansermet, CH-1211 Gen\`eve 4, Suisse}
\maketitle
\abstracts{Structure formation with topological defects is
described. The main differences from inflationary models are
highlighted. The results are compared with recent observations. It is
concluded that all the defect models studied so far are in disagreement
with recent observations of CMB anisotropies. Furthermore, present
observations do not support 'decoherence', a generic feature of
structure formation from topological defects.
}
\section{Introduction}
Even if the big bang is an
``irreproducible  experiment'', we want to learn from it
as much  as possible about the  physics at high energies. We have reasons to
hope that it may have left traces from energies much higher than those
reached in any astrophysical event or terrestrial
experiment. Therefore, even if it is irreproducible and hence not
as controllable as we might want, we simply cannot afford to ignore the
information it may have left.

The initial fluctuations in the cosmic matter density and geometry may
represent one such trace. In fact, presently there are two
relatively worked out ideas for cosmic initial fluctuations, both
relying on the physics at very high energies. In the first model,
cosmic initial perturbations are due to quantum fluctuations which
'freeze in' as classical fluctuations when they become super-horizon
during an {\em inflationary era}. 

The second possibility is that {\em topological defects} which may have
formed during a phase transition in the early universe have
induced structure formation. This second possibility is the topic of
this talk.

Here, a pedagogical remark may be  in order: Often these two
alternatives have been represented as 'inflation versus defects'. This
is of course not quite correct, as topological defects have nothing to
say about the flatness, the horizon and the monopole or moduli (or
whatever unwanted relicts) problems which inflation also solves. It
is, however, easy to construct inflationary models where the amplitude
of initial fluctuations is much too small to be relevant for structure
formation. Therefore, in a model, where cosmic structure is due to
topological defects, one needs either inflation prior to defect
formation or another mechanism to solve the flatness, horizon  and relict
problems.

The reminder of this talk is organized as follows: In Section~2 I give
a short overview on the formation of topological defects during
cosmological phase transitions.
In Section~3 I discuss the problem of structure formation
with topological defects. I will first describe  some generic insights
and then discuss results for specific models. Conclusions are
presented in Section~4.

\section{Topological defects}
During adiabatic expansion the universe  cools down from a
very hot initial state. It is natural to expect that the cosmic plasma
undergoes several symmetry breaking phase transitions. In the
process of such a transition an initial symmetry group $G$ is broken
down to a subgroup $H$.  Depending on the topology of the vacuum manifold
$\cal M$, which generically is topologically equivalent to the
homogeneous space $G/H$, topological defects may form.

 This is  described by an order parameter or  Higgs field, $\phi$,
with a temperature dependent
effective potential. The field values which minimize the potential
form the vacuum manifold $\cal M$. After the phase transition the field will
assume different values in $\cal M$ in different positions of physical
space, which are uncorrelated if, {\it e.g.} the spatial separation is
larger than the present particle horizon, $l_H\sim t$. If the topology of
the vacuum manifold is non-trivial, the Kibble mechanism~\cite{Kib}
generically leads to the formation of topological defects: the field
$\phi$  may vary in space in such a
way that there are points, where $\phi$ has to leave the vacuum
manifold by continuity reasons and assume values with higher potential
energy. Such points have to form a connected sub-manifold of
spacetime.

 For example if $\cal M$ is not
connected, $\pi_0(\MM) \neq \{ 0\}$, in different positions $\phi$ can
assume  values which belong to disconnected parts of $\MM$ and
therefore is has to leave the vacuum manifold somewhere in
between . The sub-manifold of higher energy is in this case three
dimensional in spacetime and is called  {\bf domain wall}. (Domain
walls from high energy phase transitions are disastrous for
cosmology.) Similarly, a non simply connected vacuum manifold,
$\pi_1(\MM) \neq  \{ 0\}$, leads to the formation of two dimensional
defects, {\bf cosmic strings}. Domain walls and cosmic strings are
either infinite or closed. If $\MM$ contains non shrinkable two
spheres, $\pi_2(\MM) \neq  \{ 0\}$, one dimensional defects, {\bf monopoles}
form. Finally, if  $\pi_2(\MM) \neq  \{ 0\}$, zero dimensional {\bf
textures} appear, which are events of higher energy. By Derrick's
theorem one can argue that a scalar field configuration with
non-trivial $\pi_3$ winding number ({\em i.e.} a texture knot)
contracts and eventually unwinds producing a space-time 'point' of
higher energy. A summary of this is given in Table~1; more details can
be found in Refs.~\cite{Rev,VS}.

\begin{table}[ht]
\begin{center}
\begin{tabular}{|l|l|l|l|} \hline
\multicolumn{3}{|c|}{ Homotopy $\pi_n$, ~~~
	dimension in spacetime $=d=4-1-n$} & appearance \\ \hline 
$\pi_0({\cal M}) \neq 0$, ~ $\cal M$ is disconnected& walls
	& $d=3$ & sheets in space \\  \hline
%$\cal M$ is disconnected &&&\\ \hline 
$\pi_1({\cal M}) \neq 0$ $\cal M$ contains  non shrinkable circles 
        & strings 
	& $d=2$ & lines in space \\   \hline
% non shrinkable circles&&&\\ \hline
$\pi_2({\cal M}) \neq 0$ $\cal M$ contains  non shrinkable 2-spheres 
	&monopoles 
	& $d=1$ & points in space \\  \hline
% non shrinkable 2-spheres&&&\\ \hline
$\pi_3({\cal M}) \neq 0$ $\cal M$ contains non shrinkable 3-spheres
	&texture
	& $d=0$ & events in spacetime \\  \hline
%non shrinkable 3-spheres&&&\\ \hline
\end{tabular}
\end{center}
\caption{Topological defects in four dimensional spacetime.
\label{T1}}
\end{table}

\begin{figure}[ht]
%\rule{5cm}{0.2mm}\hfill\rule{5cm}{0.2mm}
%\vskip 2.5cm
%\rule{5cm}{0.2mm}\hfill\rule{5cm}{0.2mm}
\centerline{\psfig{file=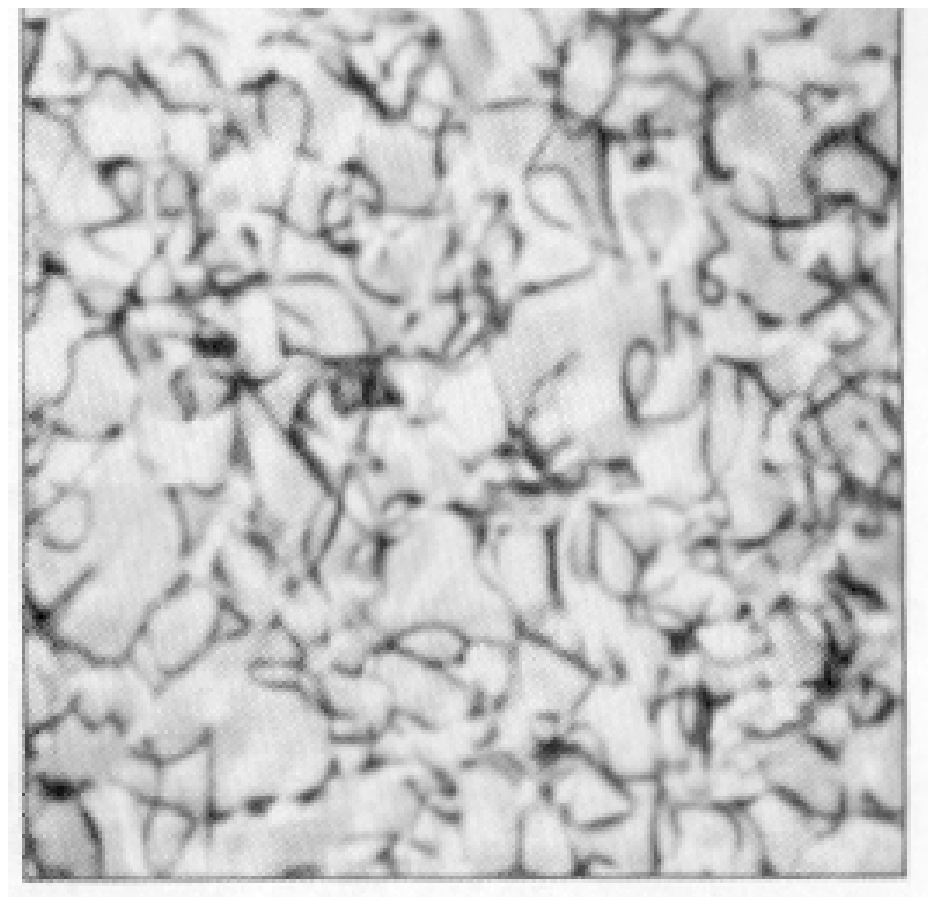,width=1.5in}
	\psfig{file=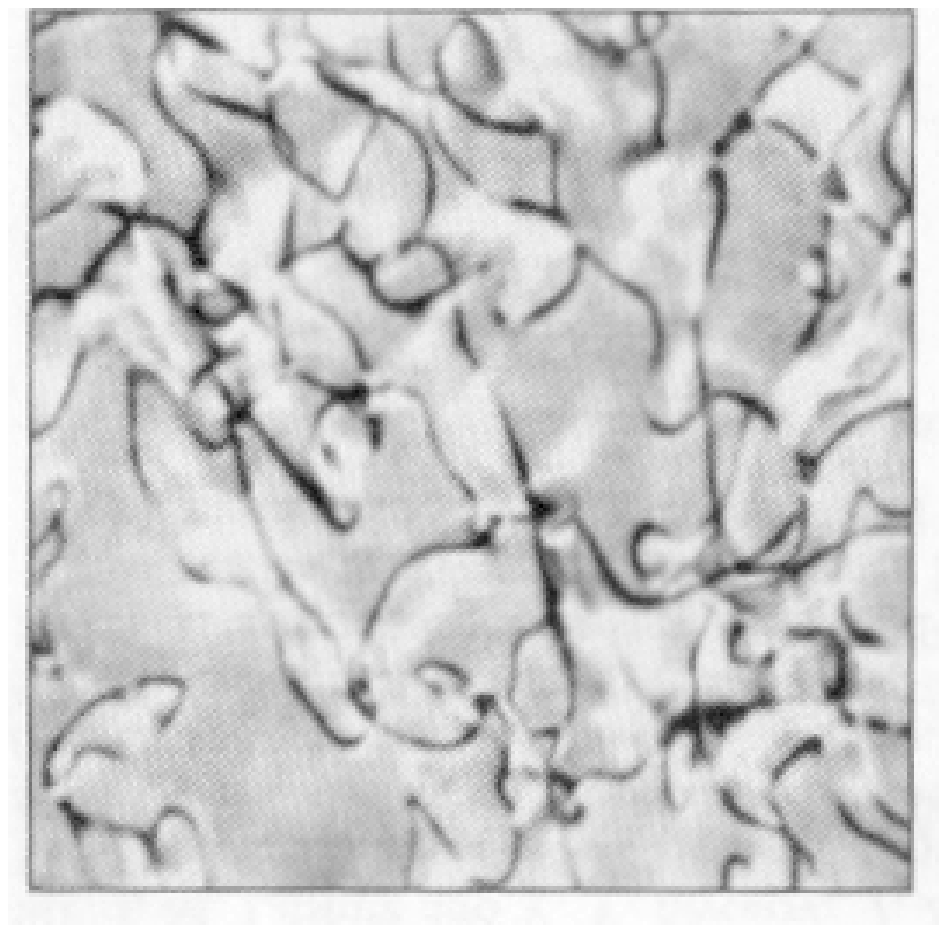,width=1.5in}
       \psfig{file=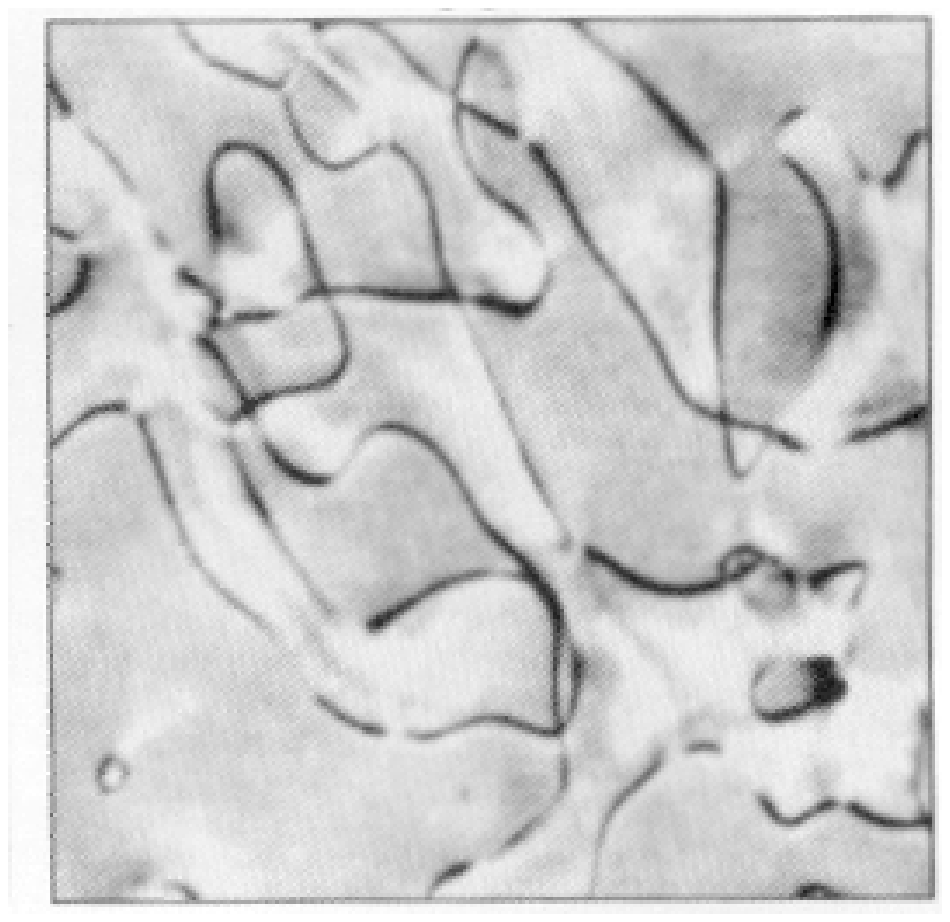,width=1.5in}
      \psfig{file=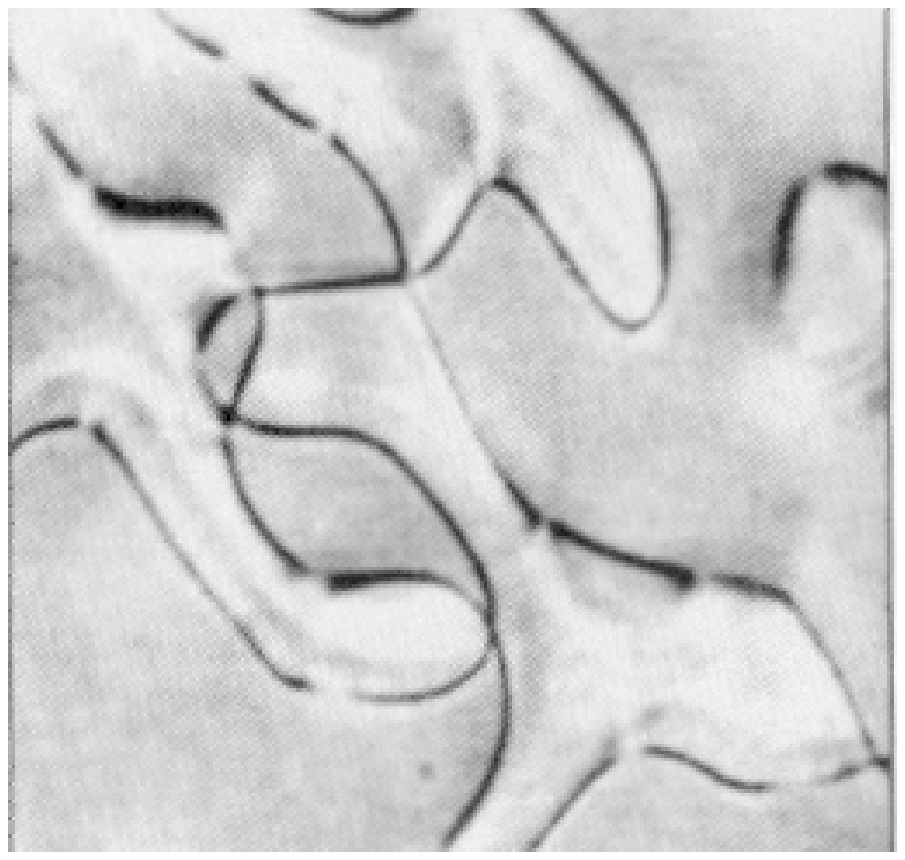,width=1.5in}}
\caption{A scaling sequence of a string network in a nematic liquid
crystal after the isotropic - nematic phase transition. Time runs from
left to right. (From 
Ref.~\protect\cite{CDTY}.)  \label{liqx}}
\end{figure}

Topological defects are also very well known
in solid state physics. For example the vortex lines in type~II super
conductors are nothing else than cosmic strings. Also in liquid
crystals~\cite{CDTY} (see Fig.~\ref{liqx}) or super fluid Helium~\cite{He} a
variety of topological defects form during symmetry breaking phase
transitions. 

 The defects are called {\bf local},
if a gauge symmetry is broken and {\bf global} if they emerge from
global  symmetry breaking. In the case of local defects, gradients in
the scalar field are 'compensated' by the gauge field and the energy
density of the defect is confined to the defect manifold with very
small transverse dimension of the order of the symmetry breaking scale. 
Soon after formation, local defects therefore seize to interact over
distances larger than the inverse symmetry breaking scale. 

The energy density of global defects is dominated by gradient energy and
hence of the order of $\rho_{\rm defect } \sim T_c^2/t^2$ where $T_c$ is
the symmetry breaking temperature and $t$ is the horizon scale, the typical
scale over which the scalar field varies. As the energy density of the
cosmic fluid also decays like $1/t^2$, global defects always
scale\footnote{Up to logarithmic corrections to the scaling law which
are especially important in the case of global cosmic strings.}
and lead to fluctuations with a typical amplitude of
\be \rho_{\rm defect }/\rho \sim 4\pi G T_c^2 = \ep ~. \label{amp}\ee

In the case of local defects only cosmic strings scale and
obey~(\ref{amp}).  Local
monopoles soon come to dominate the cosmic energy density and are
therefore ruled out from observations. Local texture die out.
To be relevant for structure formation, the defects have to induce
scaling fluctuations with an amplitude $\ep \sim 10^{-5}$ with implies 
$$T_c \sim 10^{16} {\rm GeV},$$ 
a grand unified energy scale. Topological defects
which form at lower temperature are of no relevance for structure
formation\footnote{With a possible exception of 'soft domain
walls', see {\em e.g.}~\protect\cite{Schramm} or the contribution of
M. Bucher to these proceedings.}.

\section{Structure formation with topological defects}
We discuss especially the differences of structure formation with
topological defects from inflationary initial perturbations. I
first  highlight some very generic features, then we discuss
 results for specific models.

\subsection{Generics}
The large scale fluctuations in the cosmic microwave background (CMB)
are of the same order as the deviation of the cosmic metric from a
Friedmann metric. Since these fluctuations are small, linear
perturbation theory is justified. For a cosmic fluid consisting of
radiation, massless neutrinos, baryons, cold dark
matter, possibly hot dark matter and/or a cosmological constant, we
obtain linear perturbation
equations (in Fourier space). For each wave vector $\bf k$ they are of
the form 
\be 
  DX = {\cal S}~, \label{diff}
\ee
where $X$ is a long vector describing all the random perturbation variables,
$D$ is a deterministic linear first order differential  operator and
${\cal S}$ is a random  
source term which consists of linear combinations of the energy
momentum tensor of the defect network. More details can be found, {\em e.g.} 
in Ref.~\cite{DKM}.

For inflationary perturbations ${\cal S}=0$ and the solutions are
determined entirely by the random initial conditions, $X(\bk,t_{\rm
in})$.  For most inflationary models  $X(\bk,t_{\rm in})$ is a set of
Gaussian random variables and hence their statistical properties are
entirely determined by the spectra $\cal P$ (the Fourier transforms of the two
point  functions),
\be
	\langle X_i((t_{\rm in},\bk)\ X_j^*((t_{\rm in},\bk')\rangle
	\equiv {\cal P}_{ij}(\bk)\de(\bk-\bk')~. \label{ininf}
\ee  
Here the Dirac delta is a consequence of statistical homogeneity which
we want to assume for the random process leading to the initial
perturbations. 

Be $A_i(k,t)$ the solution with initial condition   
$X_j(k,t_{\rm in})=\de_{ij}$. 
The spectra of the solution with initial 'spectrum' given by
Eq.~(\ref{ininf}) is then just
\[
 \langle X_i((t_0,\bk)\ X_j^*((t_0,\bk')\rangle =A_i(k,t_0)A_j^*(k,t_0)
 {\cal P}_{ij}(k)\de(\bk-\bk')~. 
\]
Tehrefore,  if $A_i$ is oscillating, {\em e.g.}
as a function of $kt$ so will $ \langle |X_i|^2\rangle$. This leads to
a very important feature in the CMB anisotropy spectrum, the acoustic
peaks: Prior to recombination, due to radiation pressure the
photon/baryon  plasma undergoes acoustic oscillations on subhorizon
scales. At recombination the photons become suddenly free and 'stream'
into our antennas without further interaction. Since the acoustic
oscillations of a given wave number $k$ are all in phase, the have a
fixed amplitude at decoupling. This phenomenon imprints itself in the
CMB anisotropy spectrum as a series of peaks. On very small scales,
the finite thickness of the recombination shell and free streaming
have to be taken into account which leads to an exponential damping of
the peaks (Silk damping). As we shall see below, the acoustic peaks
are very characteristic of inflationary perturbations.

If the source term $\SS$ does not vanish, the situation is different.
Equation~(\ref{diff}) can be solved by
means of a Green's function (kernel), $\GG(t,t')$, in the form
\be
X_j(t_0,\bk) =\int_{t_{in}}^{t_0}\! dt\GG_{jl}(t_0,t,\bk)\SS_l(t,\bk)~.
\label{Gsol}
\ee
Power spectra or, more generally, quadratic  expectation
values of the form
$\langle X_j(t_0,\bk)X_l^*(t_0,\bk)\rangle $ are then given by
\be
\langle X_j(t_0,\bk)X_l^*(t_0,\bk)\rangle =
 \int_{t_{in}}^{t_0}\! dt \int_{t_{in}}^{t_0} \! dt' \GG_{jm}(t_0,t,\bk)
  \GG^*_{ln}(t_0,t',\bk)
 \langle\SS_m(t,\bk)\SS_n^*(t',\bk)\rangle~. \label{power}
\ee
The only information about the source random variable which we really
need in order to compute power spectra are therefore the unequal time
 two point correlators
\be
\langle\SS_m(t,\bk)\SS_n^*(t',\bk)\rangle~. \label{2point}
\ee
This nearly trivial fact has been exploited by many workers in the 
field, for the first time probably in Ref.~\cite{ACFM} where the
decoherence of models with seeds has been discovered, and later in
Refs.~\cite{PST,Aetal,KD,DS,DKM} and others.

To determines the correlators~(\ref{2point}) one has to calculate the
unequal time correlators of the defect energy momentum tensor by means
of numerical simulations. To solve the enormous problem of dynamical range,
'scaling', statistical isotropy and causality have to be used.

Seeds from global topological defects and from cosmic strings are
'scaling' in the sense that their correlation functions 
$C_{\mu\nu\rho\la}$ defined by
\bea 
\Th_{\mu\nu}(\bk,t) &=& M^2\th_{\mu\nu}(\bk,t) ~, \\
C_{\mu\nu\rho\la}(\bk,t,t') &=&\langle\th_{\mu\nu}(\bk,t)
	\th_{\rho\la}^*(\bk,t')\rangle \label{cor}
\eea
are scale free; {\em i.e.} the only  dimensional parameters in
$C_{\mu\nu\rho\la}$ are the variables  $t,t'$ and $\bk$
themselves. Here the energy scale $M$ corresponds to the symmetry
breaking scale. One can set $M=T_c$. Up to a
certain number of dimensionless functions $F_n$ of $z=k\sqrt{tt'}$ and
$r=t/t'$, the correlation functions are then determined by the requirement
of statistical isotropy, symmetries and by their dimension. Causality
requires the 
functions $F_n$ to be analytic in $z^2$. A more detailed investigation
of these arguments and their consequences is presented in
Ref.~\cite{DK}. There it is shown that statistical isotropy and energy
momentum conservation reduce the correlators ~(\ref{cor}) for global
defects to five such functions $F_1$ to $F_5$. Since cosmic strings
loose energy by gravitational radiation, which is crucial to ensure
scaling, in this case 14 functions $F_n$ are needed to fully describe
the correlators. However, numerical simulations show that for cosmic
strings the density-density correlator is significantly larger than all the  
other components of $C_{\mu\nu\rho\la}$ which again simplifies the
problem~\cite{MB}. 

\begin{figure}[ht]
\centerline{\epsfig{figure=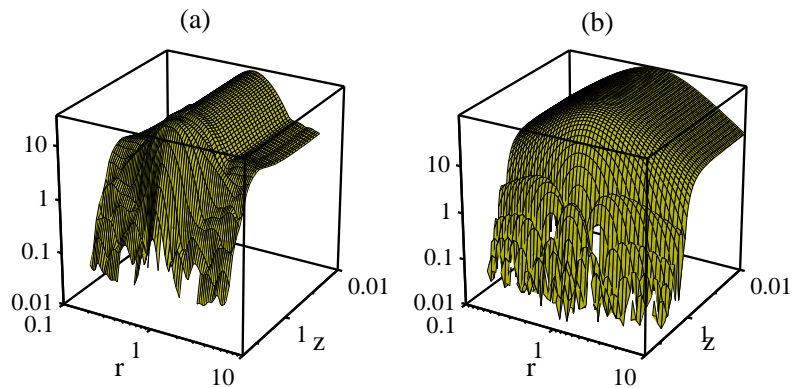,width=10.5cm}}
\caption{\label{corr} A two point correlation function of scalar perturbations 
% $C_{11}(z,r) =
% k^4\sqrt{tt'}\langle\Phi_s(\bk,t)\Phi_s^*(\bk,t')\rangle$ 
 is shown. Panel (a) represents the result from numerical simulations of
 the texture model; panel (b) shows the large-$N$ limit.  For
 fixed $r$ the correlator is constant for $z<1$ and then decays. Note
 also the symmetry under $r\ra 1/r$ (figure from Ref.~\protect\cite{DKM}).}
\end{figure}

Since analytic functions generically
are  constant for small arguments $z^2\ll 1$, $F_n(0,r)$ actually
determines $F_n$ for all values of $k$ with $z=k\sqrt{tt'}\lsim
0.5$. Furthermore, the correlation functions decay inside the horizon
and we can safely set them to zero for $z\gsim 40$ where they have
decayed by about two orders of magnitude. In Fig.~\ref{corr} I show
one of these functions for global $O(4)$-texture (a) and for the 
large $N$ limit of global $O(N)$ models\cite{DKM} (b).

For the induced perturbations in the cosmic fluids, the presence of
of this source term has several important consequences. First of all,
as is clear from Eqn.~(\ref{Gsol}), the randomness of the source term
enters at all times (as long as the source term is
non-zero). Therefore, fluctuations of a given wave number $k$ are in
general not in phase, and the distinctive series of acoustic peaks
present in inflationary models is blurred into one 'broad hump'. This
phenomenon has been termed 'decoherence'~\cite{ACFM}. A key ingredient
for decoherence to happen is the non-linearity of the time evolution
of the source term\footnote{If the source term would evolve linearly it
could just be added to the components of $X$ and we would obtain a
new, somewhat longer linear system of equations where again randomness
can enter only via the initial conditions.}. Even though
time evolution is deterministic, different Fourier modes mix due to
non-linearity, and the
randomness in one mode 'sweeps' into the other modes. In the case of
topological defects, $\SS $ is given by linear combinations of the 
defect stress energy tensor, $\theta_{\mu\nu}$,  quadratic in the
defect field, which itself
obeys non-linear evolution equations. Only in the large $N$ limit, the
evolution of the 'defect field' becomes linear and decoherence is much
weaker. The non-linearity of the source evolution also leads to the
non Gaussianity of defect models. Even if the initial field
configuration would be Gaussian (which it usually is not due to
non-linear constraints), the non-linear time evolution renders the
source term and therefore also the perturbations highly non Gaussian.

In Table~2 we highlight the similarities and differences of
inflationary and defect models of structure formation.

\begin{table}[htb]
\begin{center}
\begin{tabular}{||l| l||}
\hline\hline
 & \\
{\Large\bf Inflationary models} & {\Large\bf Topological defects}\\
\hline
\multicolumn{2}{||c||}{\large\bf Similarities} \\ \hline
\multicolumn{2}{||l||}{$\bullet$ Cosmic structure formation is due to
	gravitational instability of small 'initial' } \\ 
\multicolumn{2}{||l||}{ ~fluctuations. $\rightarrow$ Gravitational 
	 perturbation theory can be applied.} \\
\multicolumn{2}{||l||}{$\bullet$ GUT scale physics is involved in
	generating initial fluctuations.}\\
\multicolumn{2}{||l||}{$\bullet$ The only relevant 'large scale' is
	the horizon scale. $\rightarrow$ Harrison-Zel'dovich spectrum.} \\
\hline
\multicolumn{2}{||c||}{\large\bf Differences} \\ \hline
$\bullet$The  fluctuation amplitude depends on&$\bullet$The amplitude
of fluctuations is fixed\\
~details of the inflaton potential, fine tuning. &~by the
symmetry breaking scale.\\
$\bullet$Homogeneous perturbations
 (passive). &$\bullet$Inhomogeneous  perturbations (active).\\
$\bullet$Vector perturbations decay and become
&$\bullet$Vector perturbations are sourced on  \\
 ~~irrelevant. & ~large scales and are typically of the\\
 & ~ same order as 
scalar perturbations. \\
$\bullet$Tensor perturbations can have arbitrary &
$\bullet$Scalar, vector and tensor modes are \\
~amplitudes. & ~generically of the same order.\\
$\bullet$Perturbations are usually adiabatic. & $\bullet$
Perturbations are  isocurvature.\\
$\bullet$Perturbations are usually Gaussian. & $\bullet$ Perturbations
are non Gaussian.\\
$\bullet$For given initial perturbations, the&$\bullet$The
 source evolution is  non-linear at\\
~problem is linear. &~all times.\\ 
$\bullet$Randomness enters only via the  &$\bullet$Randomness enters
at all times due\\~initial conditions. &~to the mixing of scales\\
$\bullet$The phases of perturbations at a fixed
  &$\bullet$The phases  of perturbations become\\
~  scale are coherent. &~  incoherent, {\bf decoherence}.\\
$\bullet$Super Hubble scale correlations exist.&
	$\bullet$No correlations on super Hubble scales. \\
\hline\hline
\end{tabular} \end{center}
\caption{ Similarities and and differences of inflationary
perturbations versus perturbations induced by seeds.}
\end{table}
 
\subsection{Results}
As we have seen, 
there are several important differences between defect models and
inflationary models of structure formation. First of all, defect
models generically predict scalar, vector and tensor perturbations
with comparable amplitudes at horizon scale, whereas in inflationary
models  vector perturbations are absent (they simply have decayed from
their initial values) and tensor perturbations are often significantly
smaller than scalar modes. Furthermore, inflationary perturbations are
usually adiabatic. This leads to an important cancelation in the
temperature fluctuations due to gravity, given by $\left({\De T\over
T}\right)_{\rm grav} =-2\Phi$, where $\Phi$ denotes denotes the
Newtonian potential, and the intrinsic temperature fluctuation on large
scales,  which is $\left({\De T\over
T}\right)_{\rm int} ={1\over 4}\de_{rad}={1\over 3}\de_{mat}={5\over
3}\Phi$ in the adiabatic case. The net result becomes $\left({\De T\over
T}\right)_{\rm SW} =-{1\over 3}\Phi$, the ordinary Sachs-Wolfe effect
for adiabatic perturbations~\cite{SW}. 

 Both these effects lower the
temperature fluctuations of inflationary models on very large scales
if compared to those from defect models. This leads to the result that
the amplitude of fluctuations on very large scales, the height of the
'Sachs-Wolfe plateau' is comparable to the amplitude of intermediate
scales, the acoustic peak(s). This has first been noted in
Ref.~\cite{DGS}. Furthermore, the isocurvature nature of defect models
leads to a shift of the first acoustic peak towards smaller angular
scales. For flat cosmologies the peak position is around $\ell_{\rm peak}\sim
350-450$, depending on the specific model (to be compared with
$\ell_{\rm peak} \sim 220$ for inflationary models). 

\begin{figure}[ht]
\centerline{\epsfig{figure=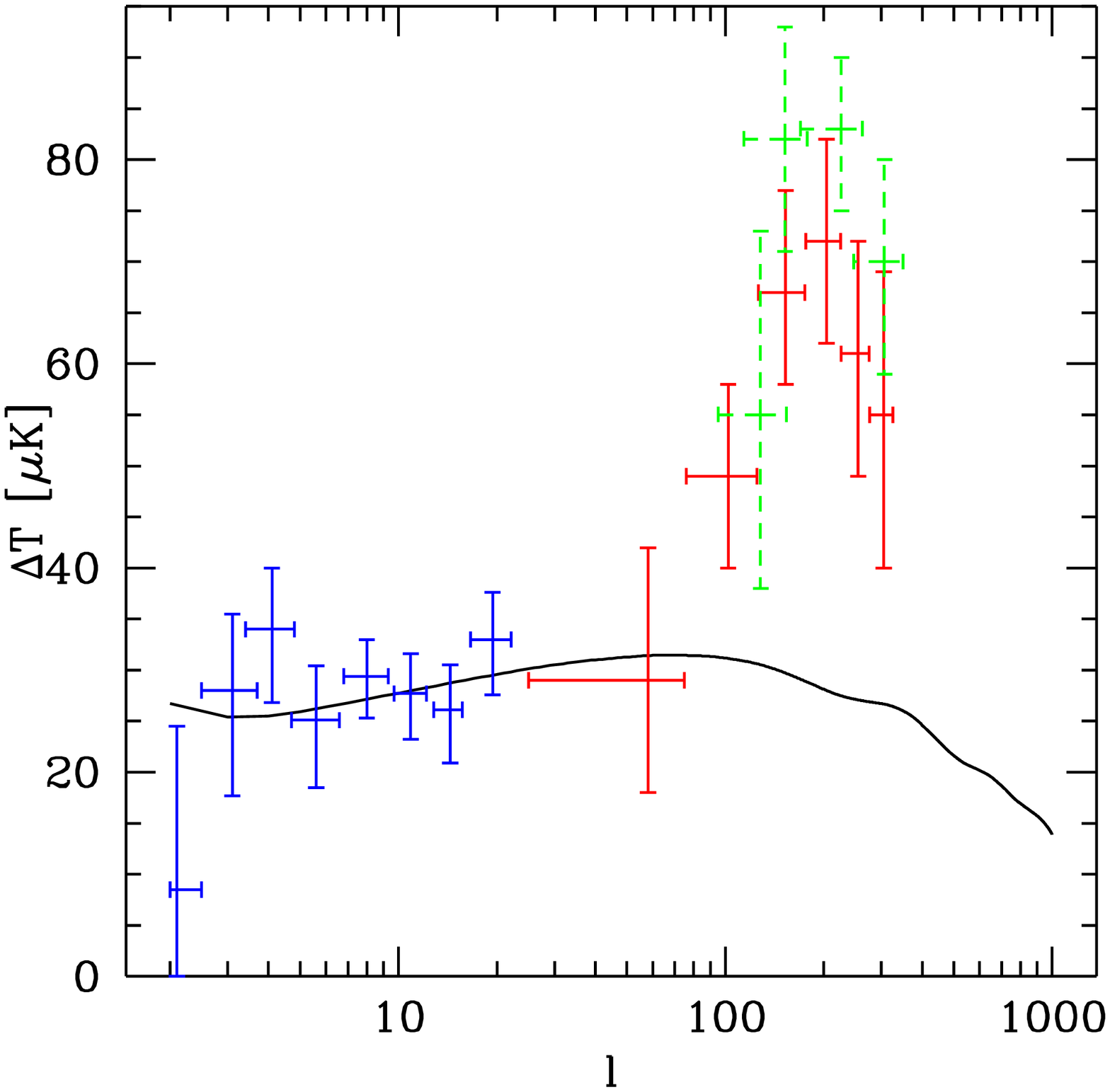,width=5.5cm}
	\epsfig{figure=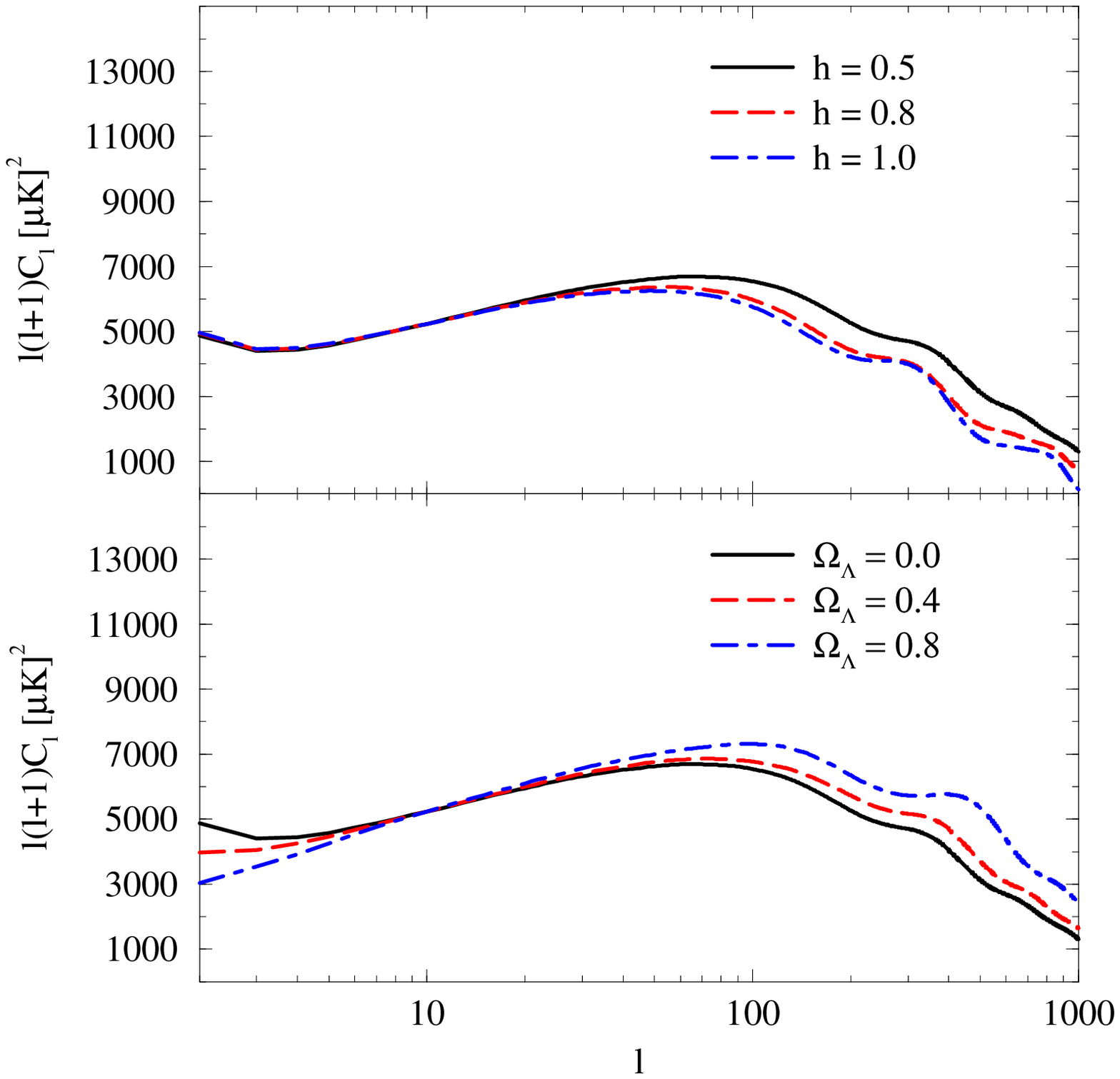,width=7cm}}
\caption{\label{texture} {\bf Left:} The CMB anisotropies obtained from global
$O(4)$ texture are compared with data, COBE~\protect\cite{Cobe} (the 8
leftmost data points),
BoomerangNA~\protect\cite{Boom} (solid crosses) and
Toco~\protect\cite{toco} (dashed crosses). The model clearly disagrees
with the Toco and Boomerang data.
{\bf Right:} The CMB anisotropies obtained from global
$O(4)$ texture are  shown for different values of cosmological
parameters. It is clear that varying the cosmological parameters cannot
save the model.}
\end{figure}

Thorough numerical
simulations from two different groups~\cite{PST,DKM} now show that CMB
anisotropies from global $O(N)$ models do not agree with present data
see Fig.~\ref{texture}). There models also require a
very high bias to fit the galaxy power spectrum and exhibit much too
low bulk flows on large scales. For example the bulk velocity on
$50h^{-1}$Mpc for the texture model is $V_{50}\sim 60$km/s whereas the
measured value is more like $V_{50}\sim 300\pm 100$km/s. 

The results for cosmic strings are somewhat more promising due to a
variety of effects. Most notably the following:
\begin{itemize}
\item The cosmic string energy density seems to be considerably higher in the
radiation era than in the matter era, therefore boosting the
fluctuations on scales which enter the horizon already in the
radiation dominated era of the universe, $ \lsim 50h^{-1}$Mpc, just
the scales where global $O(N)$ models are missing power.
\item $T_0^0=\rho$ is much larger than the other components of the
string energy momentum tensor. Being of scalar nature it induces only
scalar perturbations so that vector and tensor perturbations are
suppressed in the case of strings. 
\item Cosmic strings loose power on scales inside the horizon  by
inter-commutation and gravitational radiation. These processes are
slower than the speed of light with which global defects decay. 
Therefore, the energy momentum tensor
persists to later times, up to larger values of $kt$ than for
$O(N)$ models. This induces larger fluctuations in the dark matter.
\end{itemize}
The induced fluctuations in the dark matter may even be too large on
small scales, a problem which can be solved by introducing hot dark
matter~\cite{SMB}. The persistence of the string energy momentum tensor
induces even more decoherence~\cite{ACFM}  
than for $O(N)$ models. 

\begin{figure}[ht]
\centerline{\epsfig{figure=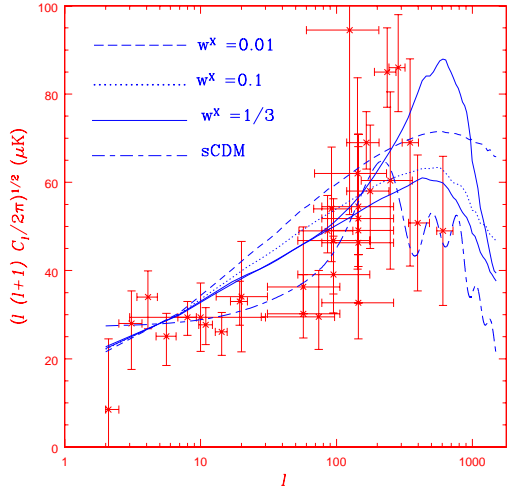,width=7cm}}
\caption{\label{joao} The CMB anisotropy spectrum for a model where
cosmic string loops 'decay' into a fluid with equation of state
$p_X=w_X\rho_X$ are shown and compared with the standard inflationary
CDM model (long-short dashed line). Clearly, the result depends very
sensitively on $w_X$. Figure taken form Ref.\protect\cite{CHM}.
}
\end{figure}

Therefore, cosmic strings may lead to one
broad 'acoustic hump' but certainly not to a series of peaks.
The precise height of the hump depends sensitively on several
unknowns, for example on how one models the string energy momentum non
conservation~\cite{CHM} and on the small scale structure of the string 
network~\cite{PV}, see, {\em e.g.}, Fig.~\ref{joao}.

Decoherence which leads to a 'smearing'
of acoustic peaks (if they are there) is one of the few features about
which all results on cosmic strings agree. The height of the
'acoustic hump' may be about two to four times the height of the
plateau at  low $\ell$. The position of the hump is not very well
defined and depends on the details of the modelling, but it is
typically at $\ell\gsim 400$ for a flat universe, which is in
disagreement with the new data shown in Fig.~\ref{texture}. The bias
factor  needed in the dark matter spectra (maybe between 2 and 5)
are still quite uncertain. Some recent work on this subject can be
found in Refs.~\cite{Aetal,BRA,AWAS,CHM}.

\section{Conclusions}
All the defect models studied in detail are in disagreement with
current observations. They exhibit no acoustic peaks (global $O(N)$ models)  
or only one broad hump on too small scales (cosmic strings). 
Decoherence, which is
inherent to the non-linear evolution of the defect source term smears
out the distinguished series of acoustic peaks expected in
inflationary models. The width of the first peak measured by the the 
Toco\cite{toco} and BoomerangNA\cite{Boom} experiments is relatively
narrow, which already clearly disfavors a model where decoherence is
important. Secondary peaks in the CMB anisotropy spectrum
will finally be a unambiguous sign for a (quasi-)linear process of
structure formation like, {\em e.g.}, inflation. 

It has been shown,
however, that linearly evolving causal scaling seeds might mimic an
inflationary CMB and dark matter power spectrum~\cite{Tu,DS}. 
Nevertheless, due to causality  they differ from inflation in the CMB 
polarization spectrum\cite{SZ}. Clearly, such seeds are not topological
defects and there is so far no convincing physical motivation to introduce
them.

\section*{Acknowledgments} It is a pleasure to acknowledge useful
discussions with Martin Kunz, Joao Magueijo and Alessandro
Melchiorri. This work is supported by the Swiss National Science
Foundation.

\end{document}